%% file: main.tex
\documentclass[a4paper,fleqn]{cas-dc}

\usepackage[numbers]{natbib}
\usepackage[ruled,vlined]{algorithm2e}
\usepackage{rotating}
\usepackage{subfigure}

\def\tsc#1{\csdef{#1}{\textsc{\lowercase{#1}}\xspace}}
\tsc{WGM}
\tsc{QE}

\usepackage{booktabs}
\usepackage{multirow}
\usepackage[ruled,vlined]{algorithm2e}
\usepackage{subfigure}
\usepackage{arydshln}

\usepackage{xcolor}

\makeatletter
\DeclareRobustCommand\onedot{\futurelet\@let@token\@onedot}
\def\@onedot{\ifx\@let@token.\else.\null\fi\xspace}

\def\eg{\emph{e.g}\onedot} 
\def\ie{\emph{i.e}\onedot} 
 
\def\etc{\emph{etc}\onedot}

\makeatother
\newcommand{\name}[0]{MABB-LKH\xspace}

\begin{document}
\let\WriteBookmarks\relax
\def\floatpagepagefraction{1}
\def\textpagefraction{.001}

\shorttitle{MAB and Backbone boost LKH Algorithm for the TSPs}    

\shortauthors{Wang et al.}  

\title [mode = title]{Multi-armed Bandit and Backbone boost Lin-Kernighan-Helsgaun Algorithm for the Traveling Salesman Problems}  

\author[]{Long Wang}
\ead{m202273734@hust.edu.cn}
\credit{Conceptualization of this study, Methodology, Software, Writing and revision}
\cormark[1]
\cortext[cor1]{The first two authors contributed equally.}
\address[]{School of Computer Science and Technology, Huazhong University of Science and Technology, China 430074}

\author[]{Jiongzhi Zheng}
\ead{jzzheng@hust.edu.cn}
\cormark[1]
\credit{Conceptualization of this study, Methodology,  Writing and revision}

\author[]{Zhengda Xiong}
\credit{Writing and revision}

\author[]{Kun He}[orcid=0000-0001-7627-4604]
\cormark[2]
\ead{brooklet60@hust.edu.cn}
\cortext[cor2]{Corresponding author.}
\credit{Conceptualization of this study, Methodology, Supervision, Writing and revision}

\begin{abstract}
The Lin-Kernighan-Helsguan (LKH) heuristic is a classic local search algorithm for the Traveling Salesman Problem (TSP). LKH introduces an $\alpha$-value to replace the traditional distance metric for evaluating the edge quality, which leads to a significant improvement. However, we observe that the $\alpha$-value does not make full use of the historical information during the search, and single guiding information often makes LKH hard to escape from some local optima. To address the above issues, we propose a novel way to extract backbone information during the TSP local search process, which is dynamic and can be updated once a local optimal solution is found. We further propose to combine backbone information, $\alpha$-value, and distance to evaluate the edge quality so as to guide the search. Moreover, we abstract their different combinations to arms in a multi-armed bandit (MAB) and use an MAB model to help the algorithm select an appropriate evaluation metric dynamically. Both the backbone information and MAB can provide diverse guiding information and learn from the search history to suggest the best metric. We apply our methods to LKH and LKH-3, which is an extension version of LKH that can be used to solve about 40 variant problems of TSP and Vehicle Routing Problem (VRP). Extensive experiments show the excellent performance and generalization capability of our proposed method, significantly improving LKH for TSP and LKH-3 for two representative TSP and VRP variants, the Colored TSP (CTSP) and Capacitated VRP with Time Windows (CVRPTW).
~~~ ~~~
\end{abstract}


\maketitle

\sloppy{}

\section{Introduction}
\label{Sec-Intro}
\input{01-Intro}

\section{Problem Definition}
\label{Sec-Prob}
\input{02-Problem_definition}

\section{Related Work}
\label{Sec-RW}
\input{03-RW}


\section{\name and \name-3}
\label{Sec-MABB-LKH}
\input{05-Method}

\section{Experimental Results}
\label{Sec-Exp}
\input{06-Exp}

\section{Conclusion}
\label{Sec-con}
\input{07-Con}

\section*{Acknowledgments}
This work is supported by 
Microsoft Research Asia (100338928).


\bibliographystyle{unsrt}

\bibliography{main}

\end{document}

%% file: 01-Intro.tex

Given an undirected, complete graph, where each node represents a city, and the distance between any two cities is known, the Traveling Salesman Problem (TSP)~\cite{Dantzig1959} aims to find the shortest Hamiltonian circuit in the graph, which starts from a city, visiting each of the other cities exactly once and finally returns to the starting city. As the basic model of many routing problems~\cite{BraunB01,NazariOST18,LiMGCLSZ22,cor/routing1,cor/routing2}, TSP is a classical NP-hard combinatorial optimization problem that has a wide range of real-world applications~\cite{cor/tsp1,cor/tsp2}.



With the increase in problem scales, the computation time for exactly solving the TSP instances grows sharply. To meet the requirements of algorithm efficiency in many real-world routing problems, heuristics are the most popular and practical methods. Heuristic methods for TSP mainly include local search~\cite{Helsgaun2000} and genetic algorithms~\cite{Nagata2013}. In this paper, we mainly focus on local search methods, which are more commonly used and more suitable for TSP instances with various scales, even for instances with over a million cities\footnote{https://www.math.uwaterloo.ca/tsp/world/index.html}.

The foremost local search algorithms are represented by two families, deriving from the Lin–Kernighan (LK)~\cite{Lin1973} method and the Stem-and-Cycle (S\&C) method~\cite{glover1996ejection}, respectively. The LK heuristic is based on the famous $k$-opt local search operator~\cite{Lin1965}, which actually adjusts the solution by replacing its $k$ edges with $k$ new edges. LK uses the distance between the endpoints to evaluate the quality of the edges. 
S\&C method proposes some particular rules based on a spanning subgraph consisting of a cycle attached to a path, which can change the solutions by moves that LK cannot generate~\cite{Rego2011}.


This paper mainly focuses on the LK series algorithms, among which the Lin–Kernighan-Helsgaun (LKH) algorithm~\cite{Helsgaun2000,Helsgaun2009} is a representative one, making many achievements in the field of TSP solving. 
LKH improves LK in many aspects, including using an $\alpha$-value derived from the 1-tree structure~\cite{Held1970,Volgenant1983} (a variant of spanning tree) to replace the distance for evaluating the edges, generalizing $k$-opt that allows non-sequential moves, chain search, tour merging, \etc. These improvements and smart designs make LKH one of the state-of-the-art local search algorithms for TSP.


LKH considers that edges with smaller $\alpha$-values are more likely to be in the optimal solution. 
For each city, LKH associates a candidate set consisting of other cities that have the smallest $\alpha$-values on the connection edges. Edges in the candidate sets are also called candidate edges. During the $k$-opt process, only candidate edges can be used to replace edges in the current solution. Specifically, the $k$-opt operator first removes $k$ edges from the current solution and reconnects them using the candidate edges, trying to improve the current solution. Obviously, the algorithm performance heavily depends on the evaluation metric used to select the candidate edges, \ie, evaluate the edge quality.



In this work, we observe that the $\alpha$-value used in LKH to evaluate the edge quality is fixed during the search. The single and fixed guidance information might limit the algorithm's search flexibility, making the algorithm hard to escape from local optima in some cases. 
To this end, we propose two approaches with learning techniques and mechanisms to provide diverse and appropriate guiding information for the local search and boost the effective LKH algorithm.

First, we propose to combine three metrics, the $\alpha$-value, distance, and backbone information, for evaluating the edge quality and ordering candidate edges. 
Towards TSP, the backbone information is represented as the frequency of edge occurrence in optimal solutions, which are however blind to local search algorithms. 
To handle this issue, an intuitive idea is to extract pseudo backbone information from the high-quality local optimal solutions generated during the search process~\cite{Zhang2005}. 
For convenience, we simply use ``backbone information'' to denote the pseudo one in the rest of this paper. The backbone information indicates that edges that appeared more frequently in those local optimal solutions are of higher quality.

The backbone information has been applied to solve TSP~\cite{Zhang2005} and has also been taken into account by LKH~\cite{Helsgaun2009}. However, 
it does not really take effect in LKH, 
which might be because it does not fully utilize the historical search information. 
In this work, we use the backbone information from a very different perspective. 
Specifically, once a local optimal solution is found, the backbone information will be updated accordingly to contain information represented by edges in historical local optimal solutions. With the accumulation of backbone information, \ie, the increase in the number of iterations, 
the backbone information becomes more accurate and valuable, and we increase its weight accordingly in the evaluation metric. 
Actually, the $\alpha$-value, distance, and backbone information evaluate each edge mainly from global, local, and historical perspectives, respectively. Our method combines their advantages and makes use of their complementarity to provide diverse guidance for the algorithm.

Second, we propose to use a multi-armed bandit (MAB) to help the algorithm learn to select a reasonable combination of the three component metrics. 
%
MAB is a basic reinforcement learning model~\cite{Slivkins2019,Lattimore2020,eswa/RL}, where the agent needs to choose and pull an arm (\ie, take an action) at each decision step (\ie, state) and gains some rewards. The agent uses the rewards to update the evaluation values of the arms, which correspond to the benefit of pulling the arms and are used as a reference for selecting the arm to be pulled. MAB can be used to help heuristic algorithms select the best element among multiple candidates~\cite{Zheng0Z0LM22}. In our method, arms in the MAB correspond to different evaluation metrics on candidate edges, \ie, different combinations of the $\alpha$-value, distance, and backbone information. 
The MAB is used to help the algorithm select a promising metric for evaluating the quality of candidate edges.

We apply our two approaches to LKH and denote the resulting algorithm as \textbf{\name} (\textbf{MAB} and \textbf{B}ackbone boost \textbf{LKH}). Both the dynamic backbone information and MAB can provide diverse guiding information for the search and learn from the search history, helping the algorithm select appropriate guiding information to escape from local optima and find better results. We further apply our methods to LKH-3~\cite{lkh3}, 
an extension of LKH for solving constrained TSPs and Vehicle Routing Problems (VRPs). The resulting algorithm is called \name-3. We select two representative TSP and VRP variant problems, the Colored TSP (CTSP) and Capacitated VRP with Time Windows (CVRPTW), to evaluate the performance of \name-3. 
Extensive experiments show the excellent and generic performance of our proposed methods.

The main contributions of this work are as follows:

\begin{itemize}
\item We propose a novel way to extract backbone information from TSP local search algorithms. The information considers all edges in historical local optimal solutions and can be updated and accumulated.
\item We propose to combine backbone information, $\alpha$-value, and distance to form a new metric for evaluating the edge quality. The new metric contains global, local, and historical information, thus can improve the algorithm robustness for various instances.
\item We propose to use an MAB to help select an appropriate combination of backbone information, $\alpha$-value, and distance. Both the MAB and backbone information can learn from the search history and provide dynamic and appropriate guiding information for the algorithm.
\item We incorporate the proposed methods into the effective LKH algorithm and its extension version, LKH-3. Extensive experiments show that both algorithms can be significantly improved, indicating the excellent performance and generalization capability of our approaches. 
\end{itemize}

%% file: 02-Problem_definition.tex
In this section, we present the definition of the involved problems, 
including the Traveling Salesman Problem (TSP), the Colored TSP (CTSP), and the Capacitated Vehicle Routing Problem with Time Windows (CVRPTW).

\subsection{Traveling Salesman Problem}
Given an undirected complete graph $G=(V,E)$, $V$ is the set containing $n$ cities and $E$ contains all pairwise edges between the cities. Each edge $(i,j) \in E$ between cities $i$ and $j$ has a cost $d(i,j)$ representing the distance. TSP aims to find a Hamiltonian circuit represented by a permutation $(c_1, c_2,..., c_n)$ of cities $\{1, 2, ..., n\}$ that minimizes the total cost, \ie $d(c_n, c_1) + \sum_{i = 1}^{n-1}{d(c_i, c_{i+1})}$.

\subsection{Colored TSP}

In the CTSP, the city set $V$ is divided into $m+1$ disjoint sets: $m$ exclusive city sets $Ec_1,Ec_2,...,Ec_m$ and one shared city set $Sc$. The cities of each exclusive set $Ec_k (k=1,2, ...,m)$ must be visited by salesman $k$ and each city from the shared city set $Sc$ can be visited by any of the $m$ salesmen. City 1 (the depot) belongs to the shared set $Sc$ and is visited by all salesmen. The CTSP needs to determine $m$ routes on the graph $G = (V, E)$ for the $m$ salesmen. Route $k$ ($k \in \{1,2,...,m\}$) can be represented by sequence ($c_1^k,c_2^k,...c_{l_k}^k,c_1^k$), where $c_1^k=1$ is the depot, $l_k$ is the number of the cities in route $k$. The $m$ routes should satisfy the following constraints. First, each city except the depot can be visited exactly once. Second, the cities belonging to exclusive set $Ec_k$ should be contained in sequence ($c_1^k,c_2^k,...c_{l_k}^k,c_1^k$). The goal of the CTSP is to find $m$ routes with the minimum total traveling distance (cost), \ie, $\sum\nolimits_{k=1}^m{d(c_1^k,c_2^k)+d(c_2^k,c_3^k)+...+d(c_{l_k}^k,c_1^k)}$.



\subsection{Capacitated VRP with Time Windows}
In the CVRPTW, every city (customer) $i$ has its requirement $r_{i}$ and wishes to be served in an expected time window, \ie, $[t_{i}^{a}, t_{i}^{b}]$. The cities are visited by multiple vehicles, which depart from the depot, visit the cities, and finally return to the depot. Each city except the depot can be visited only once by only one vehicle. The vehicles have their capacity $C$, and the total requirement of the cities visited by each vehicle cannot exceed the capacity $C$. Moreover, a vehicle can wait at city $i$ before its service begins at $t_{i}^{a}$. The CVRPTW aims to decide the number of vehicles used and the routes of the vehicles to minimize the total traveling distance (cost) of the vehicles while satisfying the time windows and capacity constraints.



%% file: 03-RW.tex
In related works, we first review some learning-based methods for TSP, including end-to-end methods with deep neural networks and the combinations of learning models and traditional algorithms, then review studies using backbone information, and finally revisits the LKH and LKH-3 algorithms.

\subsection{Learning-based Methods for TSP}
Many studies attempt to use learning-based methods to solve the typical combinatorial optimization problem of TSP,  
which can be divided into two main categories. 

The first category uses deep learning models in an end-to-end manner to directly find a solution. 
Representative models include the graph neural network~\cite{Khalil2017}, the Pointer network~\cite{Vinyals2015} and its improved version of Pointerformer~\cite{JinDPH0QS023}, and employed learning mechanisms include reinforcement learning~\cite{Bello2017,htsp} and supervised learning~\cite{PratesALLV19}. 
Some studies propose to use deep learning models to learn to perform and guide the traditional local search operators, such as 2-opt~\cite{d2020learning} and $k$-opt~\cite{ma2024learning}. Recently, Ye et al.~\cite{glop} propose the GLOP method that combines non-auto-regressive neural heuristic methods for global problem segmentation and auto-regressive neural heuristic methods for local path construction. 
These studies investigate the potential of neural network models in directly solving TSP, which is a very difficult task. 
They can hardly be competitive with efficient heuristics such as LKH, especially for large-scale problems.

The second category combines learning models with traditional algorithms to boost performance,
such as the NeuroLKH~\cite{Xin2021} and VSR-LKH~\cite{Zheng2021} algorithms. NeuroLKH uses a Sparse Graph Network (SGN) with supervised learning to generate candidate edges for LKH, showing higher performance than LKH in instances with the same structure as its training instances. 
For instances with more than 6,000 cities, whose scales are significantly larger than the training ones, the performance of NeuroLKH will degrade obviously. VSR-LKH uses reinforcement learning to train a Q-value and replaces the $\alpha$-value for evaluating the edge quality, showing higher performance than LKH. Similar to the $\alpha$-value, the Q-value does not make full use of the historical information either. 


\subsection{Backbone Information for TSP}
\label{sec-RW2}

The backbone information was initially applied to (maximum) satisfiability problems~\cite{DuboisD01,Zhang2004}, and later on gradually extended to the TSP field~\cite{Zhang2005},
where the backbone information is a concept that extracted from some high-quality local optima. 
In detail, Zhang and Looks~\cite{Zhang2005} run the algorithm with fewer iterations for 30 independent times and extract backbone information from these local optimal solutions 
so as to guide the subsequent search, 
which pays extra effort to obtain prior knowledge. LKH also takes into account the usage of backbone candidate edges~\cite{Helsgaun2009}, 
but the effect is not obvious. 
One reason is that it does not fully utilize the historical search information but only uses backbone information of local optimal solutions generated in the initial iterations to help select candidate edges.

In our method, the backbone information is dynamic and contains more comprehensive historical information, \ie, considering all edges that appeared in all local optimal solutions in history. Meanwhile, we extract backbone information in every iteration, updating and using them in real-time without preprocessing and extra calculation compared with the previous method~\cite{Zhang2005}. Moreover, we combine backbone information with $\alpha$-value and distance to form a combination metric and further use an MAB to help select a promising combination. Owing to the diversity and learning mechanism, our method exhibits excellent performance and robustness.


\subsection{Revisiting the LKH and LKH-3 Algorithms}
\label{sec-LKH}
Both LKH~\cite{Helsgaun2000} and LKH-3~\cite{lkh3} can be divided into two stages. In the first stage, the algorithms select high-quality candidate edges based on the $\alpha$-value metric. The candidate edges and their ranking in the candidate sets play an important role in the algorithms because the new edges to adjust the current solution are selected sequentially from the candidate sets. In the second stage, the algorithm repeats generating an initial solution by a function called \textit{ChooseInitialTour}() and using the search operators (\ie, $k$-opt) to improve the solution to a local optimum until the stopping criterion is reached. The procedure of improving the solution to a local optimum is encapsulated in a function called \textit{LinKernighan}(), which outputs a local optimal solution that cannot be improved by the $k$-opt operator, and such a procedure is called a trial (\ie, iteration) in LKH and LKH-3.


LKH-3 is an extension of LKH for various constraint TSPs and VRPs. LKH-3 solves these problems by transforming them into the constrained TSP~\cite{Jonker1986,Rao1980}, and uses the $k$-opt method to explore the solution space. LKH-3 allows searching in the infeasible solution space and defines different violation functions for different problems to evaluate the violation extent of the given constraints. A solution is improved by $k$-opt in LKH-3 when the violation function is reduced or the violation function is unchanged while the optimization objective is reduced. A solution with zero violation values is feasible. In summary, techniques that can be used for LKH can be easily used for LKH-3.

In the following, we will introduce the key components in LKH and LKH-3, \ie, the $\alpha$-value for selecting the candidate edges and the $k$-opt operator.

\subsubsection{The $\alpha$-value and Candidate Edges}
LKH proposes the $\alpha$-value for evaluating the edges and selecting the candidate edges. The $\alpha$-value is calculated based on a 1-tree structure~\cite{Held1970}, a variant of the spanning tree. Given a graph $G= (V, E)$, for any vertex $v \in V$, we can generate a 1-tree by first constructing a spanning tree on $V\backslash \{v\}$ and then combining it with two edges from $E$ incident to $v$. The minimum 1-tree is the 1-tree with the minimum length, i.e., the total length of its edges. We denote $L(T)$ as the length of the minimum 1-tree, which is obviously a lower bound of the length of the shortest TSP tour. Moreover, we denote $L(T(i,j))$ as the length of the minimum 1-tree containing edge $(i,j)$. 
The $\alpha$-value of edge $(i,j)$ is calculated as follows.

\begin{equation}
\alpha(i,j) = L(T(i,j)) - L(T).
\end{equation}

To further enhance the performance of $\alpha$-values, LKH applies the method of adding penalties~\cite{Held1971} to vertices to obtain a tighter lower bound. Given the final $\alpha$-values of the edges, LKH and LKH-3 associate each city $i$ with a candidate set, containing five (default value) other cities with the smallest $\alpha$-values to city $i$ (sorted in ascending order of the $\alpha$-values.), and each edge between a city and its candidate city is a candidate edge.

\begin{figure}[t]
\centering
\subfigure[Sequential 3-opt]{
\includegraphics[width=0.45\columnwidth]{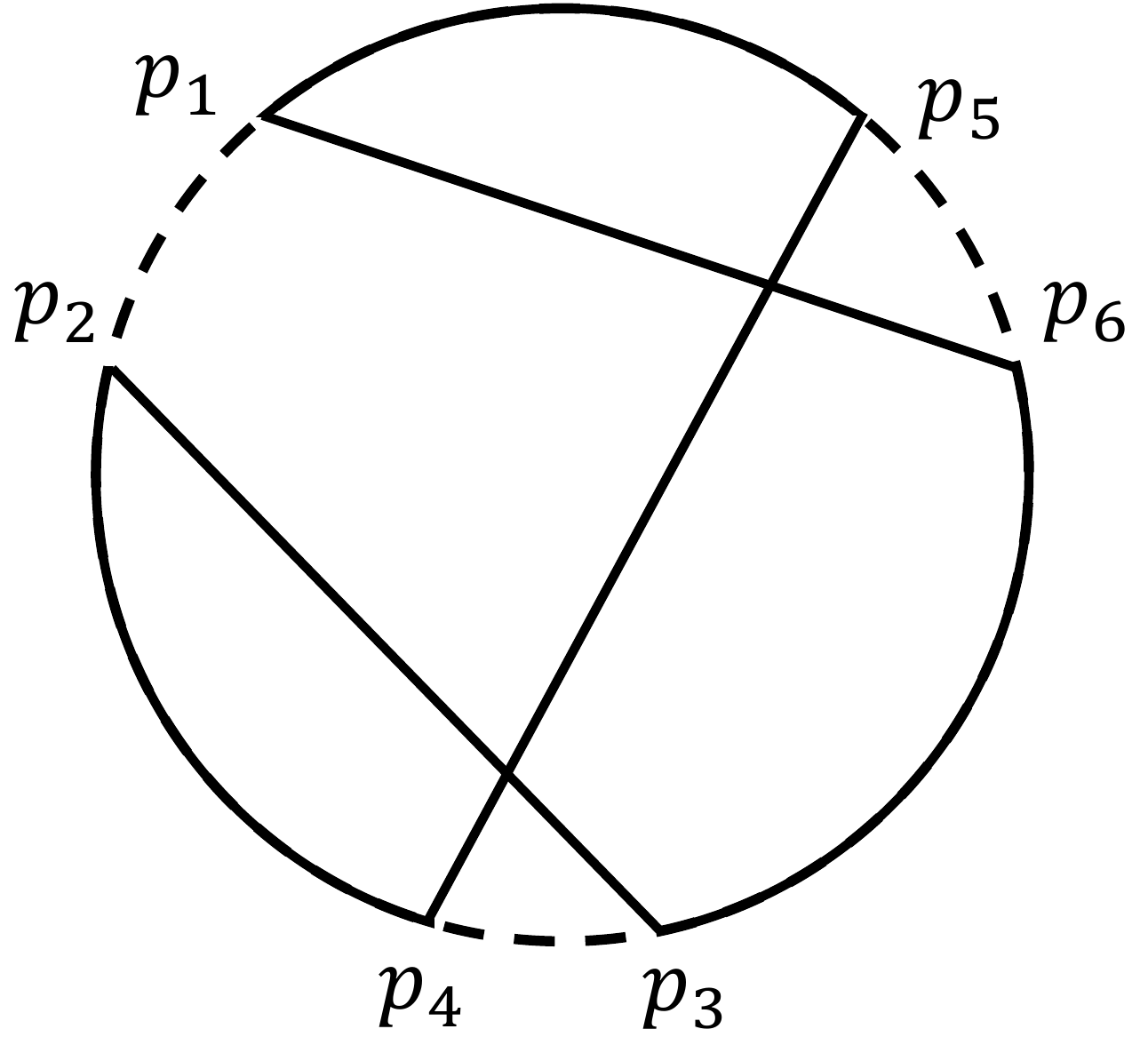}
\label{3-opt}}
\subfigure[Non-sequential 4-opt]{
\includegraphics[width=0.47\columnwidth]{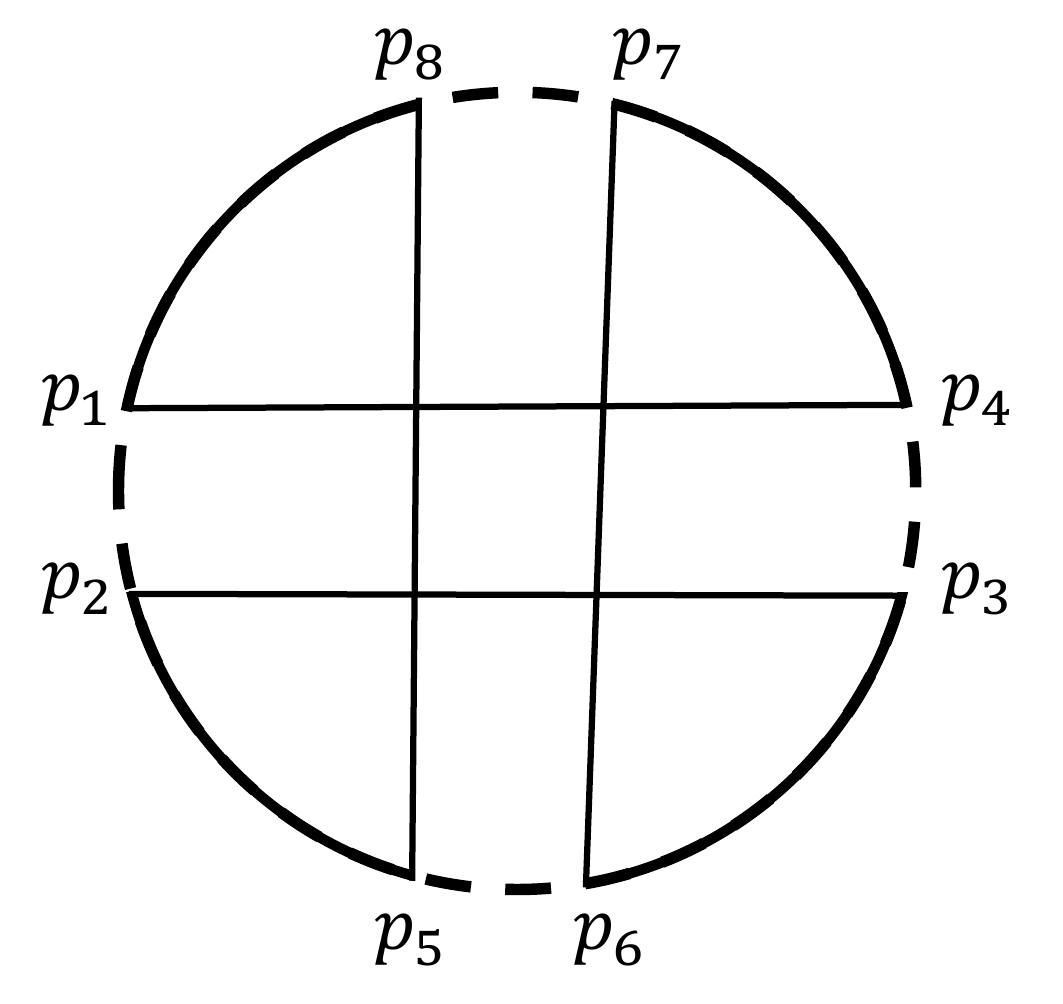}
\label{4-opt}}
\caption{Examples of sequential and non-sequential $k$-opt moves.}
\label{34-opt}
\end{figure}

\subsubsection{The $k$-opt Operator}
The $k$-opt operator in LKH and LKH-3 contains two categories, sequential and non-sequential moves, as shown in Figure \ref{34-opt}. The dashed line is the edge about to be disconnected. The sequential move starts from a starting point, \eg, $p_1$, alternatively selects the edges to be removed in the current solution (\eg, $(p_1,p_2)$, $(p_3,p_4)$ and $(p_5,p_6)$), and edges to be added sequentially from the candidate sets (\eg, $(p_2,p_3)$ and $(p_4,p_5)$), and guarantees that after selecting each edge to be removed, connecting its endpoint (\eg, $p_4$ and $p_6$) back to the starting point leads to a feasible TSP tour. Therefore, the sequential move can be stopped once an improvement is found, and the non-sequential move cannot. The non-sequential move combines two distinct infeasible $k$-opt moves to form a feasible tour, as shown in Figure \ref{4-opt}, which is a supplement of the sequential move, exploring additional search space that sequential moves cannot reach.

%% file: 05-Method.tex

The proposed \name and \name-3 algorithms improve LKH and LKH-3 from two aspects, \ie, combining backbone information, $\alpha$-value, and distance to evaluate the edge quality and selecting an appropriate combination of them by using a multi-armed bandit (MAB). The MAB model can learn during the search and adjust the ranking of the candidate cities dynamically in each iteration. Note that \name and \name-3 share a similar framework, as LKH and LKH-3 do. Therefore, this section first introduces our proposed methods that are commonly used in \name and \name-3, including how we extract and update backbone information from the historical search information, how we design the new combination metric, and the proposed MAB model, and then introduce the framework of the \name algorithm as a representative.


\subsection{Extract and Update Backbone Information}
\label{process_backbone}
In our method, the backbone information is represented by the edge frequency among the local optimal solutions, \ie, solutions outputted by the \textit{LinKernighan}() function. 
The backbone information will be updated once a local optimal solution is generated in each iteration (\ie, the trial in LKH). We define $t$ as the number of trials of the local search algorithm, and $\eta_{ij}$ as the number of times that edge $(i,j)$ appears in all the local optimal solutions in the search history. Then, the backbone information corresponding to edge $(i,j)$ is defined as:

\begin{equation}
\label{eq-b}
b_{ij} = \eta_{ij} / t.
\end{equation}

$\eta_{ij}$ is divided by $t$ which standardizes backbone information based on trials $t$. We regard that edges that appear more frequently in historical local optimal solutions should have higher quality. Note that collecting the backbone information for all edges in graph $G = (V, E)$ needs an $O(|V|^2)$ memory space. In practice, we only need to collect information for promising edges with small $\alpha$-values and lengths as LKH does, resulting in an $O(|V|)$ memory space.

\subsection{New Combined Evaluation Metric}
We define a new evaluation metric that combines the $\alpha$-value, backbone information, and distance for evaluating the edge quality and ordering cities in each candidate set to fully use their advantages corresponding to global, historical, and local perspectives, respectively. Zhang and Looks~\cite{Zhang2005} propose to combine backbone information with distance by multiplying the two metrics. In this work, we also multiply them and obtain a metric denoted as $bd$-value. The $bd$-value for an edge $(i,j)$ can be calculated as follows: 

\begin{equation}
\label{eq-bd}
bd(i,j) = (1-b_{ij})d(i,j), 
\end{equation}
which indicates that an edge with a shorter length and appearing more frequently in historical local optimal solutions will have a higher quality, measured by a lower distance. Note that the backbone information in our $bd$-value can be updated without any prior knowledge and considers all appeared edges in local optimal solutions, which is quite different from the metric in~\cite{Zhang2005}.

We furthermore combine the $\alpha$-value and $bd$-value by weighted sum with a weight factor $w$, resulting in the final new metric, $\alpha bd^w$-value. The $\alpha bd^w$-value of an edge $(i,j)$, denoted as $\alpha bd^w(i,j)$, can be calculated as follows:

\begin{equation}
\label{eq-abd}
\alpha bd^w(i,j) = w \cdot \alpha'(i,j) + (1-w) \cdot bd'(i,j),
\end{equation}
where $\alpha'(i,j) = \frac{\alpha(i,j) - \alpha_{min}}{\alpha_{max} - \alpha_{min}}$ and $bd'(i,j) = \frac{bd(i,j) - bd_{min}}{bd_{max} - bd_{min}}$ are the $\alpha$-value and $bd$-value of edge $(i,j)$ after normalization, respectively. Here $\alpha_{max}$ (resp. $bd_{max}$) and $\alpha_{min}$ (resp. $bd_{min}$) are the maximum and minimum $\alpha$-values (resp. $bd$-values) of candidate edges, respectively. This operation makes two different metrics at the same level.

Since the magnitudes of $\alpha$-value and $bd$-value might be different, and we have no idea about the best weight assignments for them in an ideal evaluation metric, we first normalize them and then use a weight factor $w$ to control their importance in the linear combination. Because sometimes $\alpha$-value is more important than $bd$-value, and sometimes the situation is the opposite. And different $w$ corresponds to different metrics. In \name, we empirically set several values of $w$ following a uniform distribution in $[0,1]$ and use an MAB model to help the algorithm select the best one.


\subsection{The MAB Model}
The MAB model is used to select an appropriate weight factor for our proposed new combined evaluation metric in each trial of \name. The metric is then used to order the candidate edges, which is quite important for the local search algorithm. Suppose the MAB has $m$ ($m > 1$) arms. We set the $i$-th arm ($i \in \{1,2,\cdots,m\}$) corresponding to a weight factor of $w_i = \frac{i-1}{m-1} \cdot \Gamma$ and also corresponding to a metric based on $\alpha bd^{w_i}$-value, where $\Gamma$ is a discount value which will decrease with the increase of trials (see details in Section~\ref{sec-usage}).

In the MAB model, each of the $m$ arms has an expected return when picked, which is hard to calculate precisely since the background problem is too complicated. Therefore, we associate each arm $i$ with an estimated value $V_i$ to approximate estimate the expected return of pulling it, which is initialized to be 0.

In the following, we first introduce the way of selecting an arm to be pulled in each step, and then introduce how to update the estimated values of the arms.

\subsubsection{Select an Arm to be Pulled}
The MAB model uses the widely-used Upper Confidence Bound (UCB) method~\cite{Hu2019,Zheng0Z0LM22} to trade-off between exploration and exploitation and selects an arm to be pulled. We denote $n_i$ as the number of times that arm $i$ has been pulled in the history, and $N = \sum_{i=1}^m{n_i}$ as the number of times calling the MAB model to pick an arm. The action (\ie, arm) selected at trial $t$ by the UCB method is:

\begin{equation}
\label{eq-UCB}
A_t = \arg\max\limits_{i}{\left(V_i + c \cdot \sqrt{\frac{\ln{N}}{n_i + 1}}\right)},
\end{equation}
where $c$ is the exploration bias parameter, also called the confidence level in the UCB method, to trade the exploitation item $V_i$ and exploration item $\sqrt{\frac{\ln{N}}{n_i + 1}}$. 


\subsubsection{Update Estimated Values}
We hope the corresponding metric selected by the MAB model can provide promising guiding information for the local search algorithm and help it find better solutions. Therefore, we use the improvement or degradation in the solution quality to calculate the reward of pulling an arm. Suppose $i$ is the arm pulled at the beginning of the current trial $t$, $R$ is the local optimal solution outputted by function \textit{LinKernighan}() at trial $t$ following the metric corresponding to arm $i$, and $R^*$ is the shortest solution found so far. We further define $L(R)$ and $L(R^*)$ as the length of solutions $R$ and $R^*$, respectively. The reward of pulling arm $i$ at trial $t$ is designed as:
\begin{equation}
\label{eq-reward}
r_t = \frac{L(R^*) - L(R)}{L(R^*) - L(T) + 1},
\end{equation}
where $L(T)$ is the lower bound of the length of the optimal solution. The numerator $L(R^*) - L(R)$ in the reward makes the reward larger for a shorter $R$, and the denominator $L(R^*) - L(T) + 1$ indicates that the closer to the optimum, the larger the reward, which is intuitive and reasonable. The extra added 1 is to avoid the situation where the denominator equals zero. 




Finally, the estimated value $V_i$ of arm $i$ pulled at trial $t$ is updated incrementally as follows.
\begin{equation}
\label{eq-update}
V_{i}^{t+1}  =  V_{i}^{t} + s \cdot (r_{t} - V_{i}^{t}),
\end{equation}
where $s$ is the step size. After updating the $V_{i}$, we sort the arms of MAB models based on $V_{i}^{t}$ dynamically so that it is a real-time model that could choose the more appropriate action.

\begin{algorithm}[!t]
\caption{\name}
\label{alg_main}
\LinesNumbered 
\KwIn{a TSP instance: $I$, the maximum number of trials: $MaxTrials$, number of trials to start using backbone information: $bs$, number of arms: $m$, step size: $s$, exploration bias parameter: $c$, weight discount factor: $\gamma$}
\KwOut{the best solution found for $I$: $R^*$}
initialize candidate sets based on methods in LKH\;
initialize length of the best solution $L(R^*) \leftarrow +\infty$\;
initialize the number of times calling MAB $N \leftarrow 0$\;
\For{$i \leftarrow 1 : m$}{initialize $V_i \leftarrow 0$, $n_i \leftarrow 0$, $w_i \leftarrow \frac{i - 1}{m - 1}$\;
}


\For{$t \leftarrow 1 : MaxTrials$}{
    $R \leftarrow$ \textit{ChooseInitialTour}()\;
    \eIf{$t \leq bs$}{
        $R \leftarrow$ \textit{LinKernighan}($I, R$)\;
    }{
        $N \leftarrow N + 1$\;
        $A_t \leftarrow \arg\max\limits_{i}{\left(V_i + c \cdot \sqrt{\frac{\ln{N}}{n_i + 1}}\right)}$\;
        $n_{A_t} \leftarrow n_{A_t} + 1$\;
        $w \leftarrow w_{A_t} \cdot \gamma^{t - bs}$\;
        sorting cities in each candidate set in ascending order according to $\alpha bd^w$-values\;
        $R \leftarrow$ \textit{LinKernighan}($I, R$)\;
        update $V_{A_t}$ according to Eq~\ref{eq-update}\;
    }
    update backbone information according to Eq~\ref{eq-b}\;
    \lIf{$L(R) < L(R^*)$}{$R^* \leftarrow R$}
}
\textbf{return} $R^*$\;
\end{algorithm}

\subsection{The Framework of \name}
\label{sec-usage}

The main framework of our \name algorithm is depicted in Algorithm~\ref{alg_main}. The algorithm first initializes the candidate sets and some important values, which will be updated during the subsequent search, including the length of the best solution $L(R^*)$, the number of times calling the MAB model $N$, as well as the estimated value $V_i$, the number of pulled times, and the initial weight factor $w_i$ of each arm $i$ (lines 1-5). Then, the algorithm repeats to search for better solutions iteratively until reaching the maximum number of trials $MaxTrials$ (lines 6-19).

In each trial, the algorithm first uses the \textit{ChooseInitialTour}() function derived from LKH to generate an initial solution $R$ (line 7), which is actually generated by adding some random perturbation based on the best solution $R^*$. Then, if the current trial $t \leq bs$, the algorithm does not use backbone information and the MAB model to adjust the candidate sets but follows the same search method of LKH and records it (lines 8-9). Actually, the first $bs$ (100 by default) trials are only used to collect backbone information. When backbone information accumulates to a basic amount (\ie, $t = bs$), the algorithm starts to use the MAB model to select an arm to be pulled $A_t$ in trial $t$ (line 12).

We argue that with the accumulation of backbone information, it will be more precise and valuable. Therefore, we use a weight discount factor $\gamma$ (0.998 by default) to increase the weight of backbone information as the number of trials increases (line 14). Actually, when $t$ is close to $bs$, $\alpha$-value still plays a major role in the evaluating metric. The backbone information will be more and more important and dominate the evaluating metric with the increase of $t$. After the algorithm determines the weight factor $w$, the evaluation metric $\alpha bd^w$-value corresponding to $w$ is then used to re-sort the candidate edges (line 15) and lead the local search function \textit{LinKernighan}() to find better solutions. Details about function \textit{LinKernighan}() are referred to Section~\ref{sec-LKH} and \cite{Helsgaun2000}.

%% file: 06-Exp.tex
\input{table-para}

\input{table-Main3}

\input{neuro_R}
\input{neuro_M}

\input{table-ctsp}

\input{table-cvrptw-solomon-25}
\input{table-cvrptw-solomon-50}
\input{table-cvrptw-solomon-100}

\input{table-cvrptw-Homberger-200}
\input{table-cvrptw-Homberger-400}
\input{table-cvrptw-Homberger-600}
\input{table-cvrptw-Homberger-800}
\input{table-cvrptw-Homberger-1000}

For experiments, we first present detailed comparison results of \name\footnote{https://github.com/JHL-HUST/MABB-LKH} 
and LKH (version 2.0.10) to evaluate the performance of our proposed new algorithm. We also compare \name with the NeuroLKH algorithm~\cite{Xin2021}, a representative learning-based algorithm for TSP. NeuroLKH combines deep learning models with LKH, using deep learning models to select candidate edges for the LKH algorithm. 
We furthermore compare \name-3 with LKH-3 in solving CTSP and CVRPTW and finally perform ablation studies to evaluate the effectiveness of the backbone information and the MAB model in \name. 


\subsection{Experimental Setup and Datasets}

\subsubsection{Experimental Setup}
\name and \name-3 were implemented in C Programming Language. The experiments were executed on a server with an AMD EPYC 7H12 CPU, running Ubuntu 18.04 Linux operating system. The tuning ranges and default values of the parameters related to backbone information and the MAB model in \name and \name-3 are shown in Table \ref{tab-para}. The parameters were tuned with an automatic configurator called SMAC3~\cite{SMAC}. 
Other parameters are consistent with the example given in the LKH\footnote{http://akira.ruc.dk/\%7Ekeld/research/LKH/} and LKH-3\footnote{http://webhotel4.ruc.dk/~keld/research/LKH-3/} open source websites. 

\subsubsection{TSP Dataset}
\label{TSP_data}
We tested the algorithms for TSP in instances from the famous TSPLIB benchmark\footnote{http://comopt.ifi.uni-heidelberg.de/software/TSPLIB95/}. 
Considering backbone information only works after 100 (\ie, $bs$) trials and the mechanism of accumulating backbone information and discounting weight, we selected all 45 symmetric TSP instances with 500 to 85,900 cities from TSPLIB as the tested benchmarks. Among them, there are 36 (resp. 7) instances with more than 1,000 (resp. 10,000) cities and two super-large instances, pla33810 and pla85900. 
Note that the number in an instance's name indicates the number of cities in the instance. The optimal solutions of all tested TSP instances are known, and the algorithms will terminate the current run and start the next one if they find the optimum. 
Following the settings of LKH, 
for each TSP instance, we set the maximum number of trials $MaxTrials$ to the number of cities and run each algorithm 10 times. Moreover, for the two super-large instances, we set $MaxTrials$ to 3,000 and run each algorithm 5 times.

\subsubsection{CTSP Datasets}

We tested \name-3 and LKH-3 for CTSP in 65 public CTSP instances that are also widely used in CTSP studies~\cite{He2021,He2021GMA}. Among the 65 instances, there are 20 small instances with 21 to 100 cities, 14 medium instances with 202 to 666 cities, and 31 large instances with 1,002 to 7,397 cities. The 65 instances are transformed from 16 TSP instances by setting different numbers of salesmen. For each CTSP instance, we set the maximum number of trials $MaxTrials$ to 10,000 and run each algorithm 10 times.



\subsubsection{CVRPTW Datasets}

We tested \name-3 and LKH-3 in two groups of widely used CVRPTW benchmarks, Solomon~\cite{Solomon87} and Homberger~\cite{Homberger1999parallel}. The Solomon benchmark contains 169 small instances, which can be divided into three sets according to the number of cities, where 57 instances have 25 cities, 56 instances have 50 cities, and 56 instances have 100 cities. The Homberger benchmark is an extension of Solomon, containing 300 instances, which can be divided into five sets of instances containing 200, 400, 600, 800, and 1,000 cities, respectively, and each set has 60 instances.

All the above CVRPTW instances with the same number of cities are divided into six groups: C1, C2, R1, R2, RC1, and RC2, each containing between 8 and 12 instances. The C1 and C2 classes have customers located in clusters, and in the R1 and R2 classes, the customers are at random positions. The RC1 and RC2 classes contain a mix of both random and clustered customers. The C2, R2, and RC2 classes have longer scheduling horizons and larger capacities than the C1, R1, and RC1 classes, meaning that each vehicle can service a larger number of customers in the former classes. For each CVRPTW instance, we also set the maximum number of trials $MaxTrials$ to 10,000 and run each algorithm 10 times.




\subsection{Comparison of \name and LKH}
\label{main_ex}
We compare our proposed \name algorithm with LKH on all the 45 tested TSP instances. 
The results are summarized in Table~\ref{tab-main}, where column \textit{Optimum} is the optimal solution, 
column \textit{Success} indicates the number of times the algorithm reaches optimum in 10 (or 5) runs, column \textit{Best} (resp. \textit{Average}) presents the best (resp. average) solutions obtained by the algorithms in 10 runs, columns \textit{Trials} and \textit{Time} indicate the average trials and running time (in seconds) of the algorithms, respectively. We split the results into two parts, containing 15 \textit{easy} instances that both LKH and \name can find the optimal solutions in each run and the other 30 \textit{hard} instances, respectively. 


From the results, one can observe that, in general, \name exhibits significantly better performance than LKH. For \textit{easy} instances, \name can usually find the optimal solutions with fewer trials and shorter running time, such as u574, dsj1000, si1032, d1655, and pla7397. For \textit{hard} instances, in some cases, such as u1060 and rl1304, \name obtains much better average results than LKH and can find the optimal solutions much more times than LKH, even for large instances like usa13509. There are some \textit{hard} instances that \name can find the optimal solutions in each of the 10 runs while LKH cannot, such as pr1002, d1291, vm1748, u2152, and fnl4461, even for large instances like rl11849. There are also some \textit{hard} instances that \name can find the optimal solutions while LKH cannot, such as rl1889, d2103, and rl5934, even for large instances like d15112. Moreover, for the super-large instances, pla33810 and pla85900, \name also shows significantly better performance than LKH.

The results indicate that \name exhibits better performance and robustness than LKH and also shows excellent performance for large instances with more than 10,000 cities. We believe that the advantages of \name over LKH are derived from the adaptive guiding information provided by the accumulated backbone information and the MAB model, which can also be demonstrated by our ablation studies (see Section~\ref{sec-ablation}).


\subsection{Comparison of \name and NeuroLKH}
NeuroLKH trains a sparse graph network with supervised learning to determine the candidate edges for LKH, which remain unchanged during the search process. 
We compare \name with two versions of NeuroLKH, NeuroLKH\_R and NeuroLKH\_M, which are trained on instances with uniformly distributed nodes and a mixture of instances with uniformly distributed nodes, clustered nodes, half uniform and half clustered nodes, respectively. 
Note that NeuroLKH only tested instances with less than 6,000 cities in its paper~\cite{Xin2021} since it costs a huge amount of resources for deep learning methods in solving large instances. 
We compare \name with NeuroLKH\_R and NeuroLKH\_M in 30 TSPLIB instances whose number of cities ranges from 500 to 6,000, 
and the detailed results are shown in Tables~\ref{neuro_R} and~\ref{neuro_M}, respectively.

The results show that \name obtains better results than NeuroLKH\_R in 8 (resp. 21) instances in terms of the best (resp. average) solutions and worse results than NeuroLKH\_R in 0 (resp. 13) instances in terms of the best (resp. average) solutions, and \name obtains better results than NeuroLKH\_M in 4 (resp. 8) instances in terms of the best (resp. average) solutions and worse results than NeuroLKH\_M in 0 (resp. 16) instances in terms of the best (resp. average) solutions. In summary, \name exhibits better performance and robustness than NeuroLKH, indicating that our learning method that allows adjusting the candidate edges during the search process is more robust than the learning method in NeuroLKH that predetermines and fixes the candidate edges, helping the LKH algorithm escaping from local optima and find better solutions.

\subsection{Comparison of \name-3 and LKH-3}



The comparison results between \name with LKH-3 in solving the 65 tested CTSP instances are shown in Table~\ref{ctsp}. The results show that \name-3 obtains better (resp. worse) results than LKH-3 in 25 (resp. 11) instances in terms of the best solutions, indicating a significant improvement in solving CTSP.

The comparison results between \name with LKH-3 in solving the three sets of Solomon CVRPTW instances are shown in Tables \ref{cvrptw-solomon-25}, \ref{cvrptw-solomon-50}, and \ref{cvrptw-solomon-100}, respectively. The comparison results between \name with LKH-3 in solving the five sets of Homberger CVRPTW instances are shown in Tables \ref{cvrptw-hoberger-200}, \ref{cvrptw-hoberger-400}, \ref{cvrptw-hoberger-600}, \ref{cvrptw-hoberger-800}, and \ref{cvrptw-hoberger-1000}, respectively. For each set of CVRPTW instances, we report the average results of the best and average solutions of all the instances in each group (\ie, C1, C2, R1, R2, RC1, RC2), as well as the average trials and running time.

Among the 48 groups of CVRPTW instances, \name-3 obtains better results than LKH-3 in 25 (resp. 26) groups in terms of the best (resp. average) solutions and worse results than LKH-3 in 11 (resp. 10) groups in terms of the best (resp. average) solutions, indicating a significant improvement.

In summary, the results in this subsection show that our proposed methods can also significantly improve the effective LKH-3 algorithm for solving TSP and VRP variants, indicating the excellent performance and generalization capability of our methods.

\subsection{Ablation Study}
\label{sec-ablation}

\begin{figure}[t]
    \centering
    \includegraphics[width=1.0\columnwidth]
    {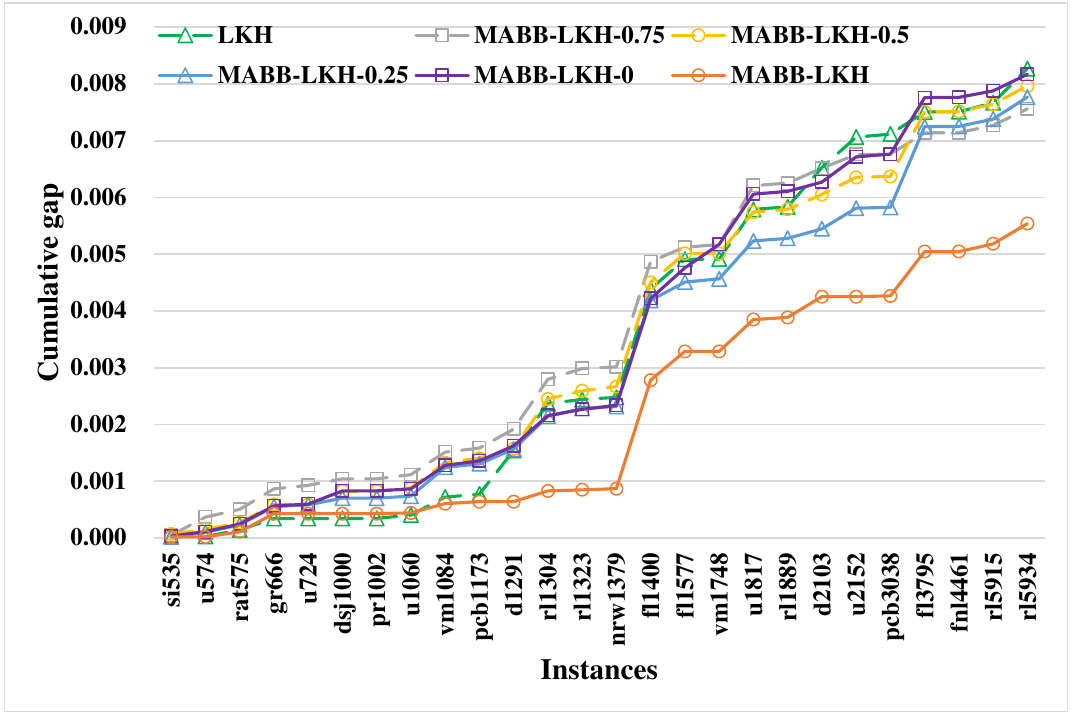}
    \caption{Comparison of \name to its variants and LKH. 
    }
    \label{fig-ablation}
\end{figure}

Finally, we perform ablation studies by comparing \name with its variants to analyze the effectiveness of our proposed methods, including the backbone information and the MAB model. Specifically, we compare \name with its four variants, as well as the LKH algorithm. The variants are denoted as MABB-LKH-$w$, where $w = \{0,0.25,0.5,0.75\}$, which uses the sole $\alpha bd^w$-value as the evaluation metric to adjust the candidate sets. In other words, MABB-LKH-$w$ is a variant of \name where the MAB model only has one arm corresponding to $w$. Actually, LKH can be regarded as MABB-LKH-1, and MABB-LKH-0 only uses the $bd$-value that multiplies backbone information and distance as the evaluation metric. The comparison results of the six algorithms are depicted in Figure~\ref{fig-ablation}, which are also expressed by the cumulative gap.

From the results, one can observe that our proposed \name algorithm significantly outperforms the algorithms with a single arm, \ie, single guiding information, including LKH, indicating that algorithms following single guidance are easy to get trapped in some local optima and our proposed MAB model can help the algorithm select appropriate guiding information and jump out of the local optima. The performance of each algorithm with a single metric is similar and not as ideal, indicating that designing a wonderful evaluation metric empirically is very hard, and our proposed MAB model suggests a way of using learning-based methods to assist local search algorithms.

Results in Figure~\ref{fig-ablation} also show that \name and its four variants perform better in solving large instances than smaller ones, as LKH is leading in solving instances with less than 1,060 cities but gets the largest cumulative gap in the end. The results indicate again that our backbone information needs accumulation to become more precise and valuable, and our method of accumulating backbone information is reasonable and effective. Moreover, MABB-LKH-0 also exhibits similar performance to other variants and LKH, indicating that without the ingenious $\alpha$-value, the backbone information can also well guide the local search algorithm to find high-quality solutions.





%% file: table-para.tex
\begin{table*}[!t]
\caption{Parameter settings of \name and \name-3.}
\centering
\footnotesize
\begin{tabular}{lllr} \toprule
Parameter & Description &Range                                 & Value \\ \hline
$bs$    &Number of trials to start using the backbone information   & $\{50, 100, 200, 500\}$                 & 100   \\
$m$     &Number of arms in the MAB model  & $\{3,4,5,6,7\}$                         & 5     \\
$s$     &Step size for updating the estimated values   & $\{0.01, 0.02, \cdots, 0.1\}$                & 0.06  \\
$c$     &Exploration bias for trade-off exploration and exploitation   & $\{0.5, 1, 2, 5, 10, 20, 50\}$          & 20    \\
$\gamma$  &Weight discount factor       & $\{0.995, 0.996, 0.997, 0.998, 0.999\}$ & 0.998 \\ \bottomrule
\end{tabular}
\label{tab-para}
\end{table*}

%% file: table-Main3.tex
\begin{table*}[!t]
\caption{Comparison results of \name and LKH on all 45 tested TSP instances. The best results appear in \bf{bold}.}
\centering
\footnotesize
\resizebox{\linewidth}{!}{
\begin{tabular}{lrrrrrrrrrrrr} \toprule
\multirow{2}{*}{Instance} & \multirow{2}{*}{Optimum} & \multicolumn{5}{c}{LKH} &                                                                       & \multicolumn{5}{c}{\name}                                         \\ \cline{3-7} \cline{9-13} 
                          &                       & Success & Best                          & Average                         & Trials  & Time(s) & & Success & Best            & Average             & Trials  & Time(s)  \\ \hline
\multicolumn{2}{l}{\textit{Easy}   instances}              &         &                               &          &                       &         &          &         &                 &                     &         &          \\
att532                    & 27686                 & 10/10  & \textbf{27686}                 & \textbf{27686.0}                   & 71.2    & 0.42   &  & 10/10  & \textbf{27686}   & \textbf{27686.0}       & 72.3    & 0.40     \\
ali535                    & 202339                & 10/10  & \textbf{202339}                 & \textbf{202339.0}                   & 4.9     & 0.12  &   & 10/10  & \textbf{202339}   & \textbf{202339.0}       & 4.9     & 0.12     \\
pa561                     & 2763                  & 10/10  & \textbf{2763}                 & \textbf{2763.0}                   & 58.5    & 0.49   &  & 10/10  & \textbf{2763}   & \textbf{2763.0}       & 41.1    & 0.39     \\
u574                      & 36905                 & 10/10  & \textbf{36905}                 & \textbf{36905.0}                   & 149.9   & 0.97  &   & 10/10  & \textbf{36905}   & \textbf{36905.0}       & 51.1    & 0.41     \\
p654                      & 34643                 & 10/10  & \textbf{34643}                 & \textbf{34643.0}                   & 22.9    & 7.78  &   & 10/10  & \textbf{34643}   & \textbf{34643.0}       & 22.9    & 7.86     \\
\bottomrule
\end{tabular}}
\label{tab-main}
\end{table*}

%% file: neuro_R.tex
\begin{table*}[!t]
\caption{Comparison results of \name and NeuroLKH\_R. The best results appear in \textbf{bold}.}
\centering
\footnotesize

\begin{tabular}{lrrrrrrrrrr}    \toprule
                              & \multicolumn{1}{c}{}       & \multicolumn{4}{c}{NeuroLKH\_R}                                                                                   & \multicolumn{1}{c}{} & \multicolumn{4}{c}{\name}                                                                          \\   \cline{3-6} \cline{8-11}
\multicolumn{1}{c}{Instances} & \multicolumn{1}{c}{BKS}    & \multicolumn{1}{r}{Best} & \multicolumn{1}{r}{Average} & \multicolumn{1}{r}{Trials} & \multicolumn{1}{r}{Time(s)} & \multicolumn{1}{r}{} & \multicolumn{1}{r}{Best} & \multicolumn{1}{r}{Average} & \multicolumn{1}{r}{Trials} & \multicolumn{1}{r}{Time(s)} \\    \hline
u574                          & 36905                      & \textbf{36905}           & \textbf{36905.0}            & 3.8                        & 0.20                        & \textbf{}            & \textbf{36905}           & \textbf{36905.0}            & 51.1                       & 0.41                        \\
rat575                        & \multicolumn{1}{r}{6773}   & \textbf{6773}            & \textbf{6773.1}             & 179.0                      & 2.93                        & \textbf{}            & \textbf{6773}            & 6773.6                      & 446.7                      & 2.49                        \\
p654                          & \multicolumn{1}{r}{34643}  & 34645                    & 34682.8                     & 548.8                      & 144.37                      & \textbf{}            & \textbf{34643}           & \textbf{34643.0}            & 22.9                       & 7.86                        \\
d657                          & \multicolumn{1}{r}{48912}  & \textbf{48912}           & 48912.5                     & 511.5                      & 7.65                        & \textbf{}            & \textbf{48912}           & \textbf{48912.0}            & 47.0                       & 0.25                        \\
\bottomrule
\end{tabular}
\label{neuro_R}
\end{table*}

%% file: neuro_M.tex
\begin{table*}[!t]
\caption{Comparison results of \name and NeuroLKH\_M. The best results appear in \textbf{bold}.}
\centering
\footnotesize
    
\begin{tabular}{lrrrrrrrrrr}    \toprule
                              & \multicolumn{1}{c}{}       & \multicolumn{4}{c}{NeuroLKH\_M}                                                                                   & \multicolumn{1}{c}{} & \multicolumn{4}{c}{\name}                                                                          \\   \cline{3-6} \cline{8-11} 
\multicolumn{1}{c}{Instances} & \multicolumn{1}{c}{BKS}    & \multicolumn{1}{r}{Best} & \multicolumn{1}{r}{Average} & \multicolumn{1}{r}{Trials} & \multicolumn{1}{r}{Time(s)} & \multicolumn{1}{r}{} & \multicolumn{1}{r}{Best} & \multicolumn{1}{r}{Average} & \multicolumn{1}{r}{Trials} & \multicolumn{1}{r}{Time(s)} \\    \hline
u574                          & 36905                      & \textbf{36905}           & \textbf{36905.0}            & 1.9                        & 0.11                        & \textbf{}            & \textbf{36905}           & \textbf{36905.0}            & 51.1                       & 0.41                        \\
rat575                        & \multicolumn{1}{r}{6773}   & \textbf{6773}            & \textbf{6773.3}             & 345.3                      & 3.73                        & \textbf{}            & \textbf{6773}            & 6773.6                      & 446.7                      & 2.49                        \\
p654                          & \multicolumn{1}{r}{34643}  & \textbf{34643}           & \textbf{34643.0}            & 7.0                        & 8.02                        & \textbf{}            & \textbf{34643}           & \textbf{34643.0}            & 22.9                       & 7.86                        \\
d657                          & \multicolumn{1}{r}{48912}  & \textbf{48912}           & \textbf{48912.0}            & 10.0                       & 0.67                        & \textbf{}            & \textbf{48912}           & \textbf{48912.0}            & 47.0                       & 0.25                        \\
\bottomrule
\end{tabular}
\label{neuro_M}
\end{table*}

%% file: table-ctsp.tex
\begin{table*}[!t]
\caption{Comparison results of \name-3 and LKH-3 in CTSP. The best results appear in \textbf{bold}.}
\centering
\footnotesize

\begin{tabular}{lrrrrrrrrr} \toprule
            & \multicolumn{4}{c}{LKH-3}                                                                      &  & \multicolumn{4}{c}{\name-3}                                                                        \\ \cline{2-5} \cline{7-10}
Instance    & \multicolumn{1}{r}{Best} & \multicolumn{1}{r}{Average} & \multicolumn{1}{r}{Trials} & Time(s)  &  & \multicolumn{1}{r}{Best} & \multicolumn{1}{r}{Average} & \multicolumn{1}{r}{Trials} & \multicolumn{1}{r}{Time(s)} \\  \hline
eil21-2     & \textbf{144918}          & \textbf{144918.0}           & 1.0                        & 0.00     &  & \textbf{726813}          & \textbf{726813.0}           & 92.0                       & 0.02                        \\
eil21-3     & \textbf{157477}          & \textbf{157477.0}           & 1.0                        & 0.00     &  & \textbf{779143}          & 780318.2                    & 833.2                      & 1.53                        \\
eil31-2     & \textbf{259356}          & \textbf{259356.0}           & 247.0                      & 0.00     &  & \textbf{759545}          & \textbf{759545.0}           & 20.0                       & 0.01                        \\
eil31-3     & \textbf{295306}          & \textbf{295306.0}           & 9.0                        & 0.00     &  & \textbf{798845}          & \textbf{798845.0}           & 7.0                        & 0.01                        \\
eil31-4     & \textbf{315964}          & \textbf{315964.0}           & 4.0                        & 0.00     &  & \textbf{144918}          & \textbf{144918.0}           & 1.0                        & 0.00                        \\
eil41-2     & \textbf{346237}          & \textbf{346237.0}           & 13.0                       & 0.00     &  & \textbf{157477}          & \textbf{157477.0}           & 1.0                        & 0.00                        \\
\bottomrule
\end{tabular}
\label{ctsp}
\end{table*}

%% file: table-cvrptw-solomon-25.tex
\begin{table*}[!t]
\caption{Comparison results of \name-3 and LKH-3 in Solomom CVRPTW dataset with 25 cities. The best results appear in \textbf{bold}.}
\centering
\footnotesize

\begin{tabular}{lrrrrrrrrr} \toprule
\multirow{2}{*}{Instance group}           & \multicolumn{4}{c}{LKH-3}                                                                                         & \multicolumn{1}{c}{} & \multicolumn{4}{c}{\name-3}                                                                          \\   \cline{2-5} \cline{7-10}
 & \multicolumn{1}{r}{Best} & \multicolumn{1}{r}{Average} & \multicolumn{1}{r}{Trials} & \multicolumn{1}{r}{Time(s)} & \multicolumn{1}{r}{} & \multicolumn{1}{r}{Best} & \multicolumn{1}{r}{Average} & \multicolumn{1}{r}{Trials} & \multicolumn{1}{r}{Time(s)} \\    \hline
C1         & \textbf{1905.89}         & \textbf{1905.89}            & 4.4                        & 0.00                        &                      & \textbf{1905.89}         & \textbf{1905.89}            & 4.4                        & 0.00                        \\
C2         & \textbf{2144.50}         & \textbf{2144.50}            & 127.0                      & 0.07                        &                      & \textbf{2144.50}         & \textbf{2144.50}            & 127.0                      & 0.12                        \\
R1         & \textbf{5004.25}         & \textbf{5004.25}            & 11.3                       & 0.00                        &                      & \textbf{5004.25}         & \textbf{5004.25}            & 11.3                       & 0.00                        \\
R2         & \textbf{3821.45}         & \textbf{3821.45}            & 144.9                      & 0.04                        &                      & \textbf{3821.45}         & \textbf{3821.45}            & 121.7                      & 0.10                        \\
RC1        & \textbf{3502.38}         & \textbf{3502.38}            & 19.4                       & 0.00                        &                      & \textbf{3502.38}         & \textbf{3502.38}            & 19.4                       & 0.00                        \\
\bottomrule                     
\end{tabular}
\label{cvrptw-solomon-25}
\end{table*}

%% file: table-cvrptw-solomon-50.tex
\begin{table*}[!t]
\caption{Comparison results of \name-3 and LKH-3 in Solomom CVRPTW dataset with 50 cities. The best results appear in \textbf{bold}.}
\centering
\footnotesize

\begin{tabular}{lrrrrrrrrr} \toprule
\multirow{2}{*}{Instance group}           & \multicolumn{4}{c}{LKH-3}                                                                                         & \multicolumn{1}{c}{} & \multicolumn{4}{c}{\name-3}                                                                          \\   \cline{2-5} \cline{7-10}
 & \multicolumn{1}{r}{Best} & \multicolumn{1}{r}{Average} & \multicolumn{1}{r}{Trials} & \multicolumn{1}{r}{Time(s)} & \multicolumn{1}{r}{} & \multicolumn{1}{r}{Best} & \multicolumn{1}{r}{Average} & \multicolumn{1}{r}{Trials} & \multicolumn{1}{r}{Time(s)} \\    \hline
C1         & \textbf{3616.89}         & \textbf{3616.89}            & 8.0                        & 0.00                        &                      & \textbf{3616.89}         & \textbf{3616.89}            & 8.0                        & 0.00                        \\
C2         & \textbf{3575.00}         & \textbf{3575.00}            & 14.8                       & 0.00                        &                      & \textbf{3575.00}         & \textbf{3575.00}            & 14.8                       & 0.00                        \\
R1         & 7665.33                  & 7670.59                     & 257.1                      & 0.29                        &                      & \textbf{7661.33}         & \textbf{7668.14}            & 232.7                      & 0.23                        \\
R2         & \textbf{6153.82}         & 6161.30                     & 383.7                      & 0.19                        &                      & \textbf{6153.82}         & \textbf{6159.00}            & 359.7                      & 0.23                        \\
RC1        & \textbf{7303.13}         & \textbf{7303.13}            & 54.5                       & 0.02                        &                      & \textbf{7303.13}         & \textbf{7303.13}            & 55.9                       & 0.02                        \\
\bottomrule                  
\end{tabular}
\label{cvrptw-solomon-50}
\end{table*}

%% file: table-cvrptw-solomon-100.tex
\begin{table*}[!t]
\caption{Comparison results of \name-3 and LKH-3 in Solomom CVRPTW dataset with 100 cities. The best results appear in \textbf{bold}.}
\centering
\footnotesize

\begin{tabular}{lrrrrrrrrr} \toprule
\multirow{2}{*}{Instance group}            & \multicolumn{4}{c}{LKH-3}                                                                                         & \multicolumn{1}{c}{} & \multicolumn{4}{c}{\name-3}                                                                          \\  \cline{2-5} \cline{7-10}
 & \multicolumn{1}{r}{Best} & \multicolumn{1}{r}{Average} & \multicolumn{1}{r}{Trials} & \multicolumn{1}{r}{Time(s)} & \multicolumn{1}{r}{} & \multicolumn{1}{r}{Best} & \multicolumn{1}{r}{Average} & \multicolumn{1}{r}{Trials} & \multicolumn{1}{r}{Time(s)} \\   \hline
C1          & \textbf{8267.00}         & \textbf{8267.00}            & 43.8                       & 0.02                        &                      & \textbf{8267.00}         & \textbf{8267.00}            & 44.9                       & 0.03                        \\
C2          & \textbf{5873.75}         & \textbf{5873.75}            & 14.4                       & 0.01                        &                      & \textbf{5873.75}         & \textbf{5873.75}            & 14.4                       & 0.01                        \\
R1          & 11750.58                 & \textbf{11779.80}           & 5676.4                     & 12.00                       &                      & \textbf{11737.58}        & 11783.01                    & 5782.7                     & 18.77                       \\
R2          & 8732.45                  & 8749.45                     & 5710.5                     & 6.25                        &                      & \textbf{8725.27}         & \textbf{8746.58}            & 4156.7                     & 7.96                        \\
RC1         & 13355.00                 & 13414.39                    & 7045.9                     & 16.20                       &                      & \textbf{13346.75}        & \textbf{13380.29}           & 6305.0                     & 24.56                       \\
\bottomrule        
\end{tabular}
\label{cvrptw-solomon-100}
\end{table*}

%% file: table-cvrptw-Homberger-200.tex
\begin{table*}[!t]
\caption{Comparison results of \name-3 and LKH-3 in Homberger CVRPTW dataset with 200 cities. The best results appear in \textbf{bold}.}
\centering
\footnotesize

\begin{tabular}{lrrrrrrrrr} \toprule
\multirow{2}{*}{Instance group}             & \multicolumn{4}{c}{LKH-3}                                                                                         & \multicolumn{1}{c}{} & \multicolumn{4}{c}{\name-3}                                                                        \\ \cline{2-5} \cline{7-10}
 & \multicolumn{1}{r}{Best} & \multicolumn{1}{r}{Average} & \multicolumn{1}{r}{Trials} & \multicolumn{1}{r}{Time(s)} & \multicolumn{1}{r}{} & \multicolumn{1}{r}{Best} & \multicolumn{1}{r}{Average} & \multicolumn{1}{r}{Trials} & \multicolumn{1}{r}{Time(s)} \\  \hline
C1           & 2729.00                  & 2745.04                     & 5891.3                     & 29.15                       &                      & \textbf{2719.37}         & \textbf{2744.60}            & 5844.8                     & 95.75                       \\
C2           & 1833.43                  & 1834.72                     & 4123.8                     & 16.42                       &                      & \textbf{1831.96}         & \textbf{1833.74}            & 4484.9                     & 39.14                       \\
R1           & 3668.98                  & 3701.45                     & 10000.0                    & 68.44                       &                      & \textbf{3653.19}         & \textbf{3700.46}            & 9950.4                     & 241.84                      \\
R2           & 2935.00                  & 2958.81                     & 9289.5                     & 76.61                       &                      & \textbf{2932.30}         & \textbf{2955.62}            & 8810.4                     & 174.26                      \\
RC1          & 3251.59                  & 3315.74                     & 10000.0                    & 98.48                       &                      & \textbf{3233.63}         & \textbf{3301.07}            & 10000.0                    & 349.61                      \\
\bottomrule
\end{tabular}
\label{cvrptw-hoberger-200}
\end{table*}

%% file: table-cvrptw-Homberger-400.tex
\begin{table*}[!t]
\caption{Comparison results of \name-3 and LKH-3 in Homberger CVRPTW dataset with 400 cities. The best results appear in \textbf{bold}.}
\centering
\footnotesize

\begin{tabular}{lrrrrrrrrr} \toprule
\multirow{2}{*}{Instance group}             & \multicolumn{4}{c}{LKH-3}                                                                                         & \multicolumn{1}{c}{} & \multicolumn{4}{c}{\name-3}                                                                        \\ \cline{2-5} \cline{7-10}
 & \multicolumn{1}{r}{Best} & \multicolumn{1}{r}{Average} & \multicolumn{1}{r}{Trials} & \multicolumn{1}{r}{Time(s)} & \multicolumn{1}{r}{} & \multicolumn{1}{r}{Best} & \multicolumn{1}{r}{Average} & \multicolumn{1}{r}{Trials} & \multicolumn{1}{r}{Time(s)} \\  \hline
C1           & 7289.88                  & 7424.98                     & 7389.9                     & 141.89                      &                      & \textbf{7264.20}         & \textbf{7392.78}            & 7142.1                     & 500.20                      \\
C2           & 4017.90                  & 4120.33                     & 8965.4                     & 301.38                      &                      & \textbf{4013.89}         & \textbf{4089.79}            & 8596.4                     & 920.57                      \\
R1           & 8739.23                  & 8827.32                     & 10000.0                    & 201.27                      &                      & \textbf{8685.85}         & \textbf{8807.38}            & 10000.0                    & 842.74                      \\
R2           & 6215.85                  & 6275.86                     & 10000.0                    & 205.14                      &                      & \textbf{6196.04}         & \textbf{6262.78}            & 10000.0                    & 756.90                      \\
RC1          & 8262.62                  & 8373.54                     & 10000.0                    & 256.73                      &                      & \textbf{8183.10}         & \textbf{8360.83}            & 10000.0                    & 1022.65                     \\
\bottomrule
\end{tabular}
\label{cvrptw-hoberger-400}
\end{table*}

%% file: table-cvrptw-Homberger-600.tex
\begin{table*}[!t]
\caption{Comparison results of \name-3 and LKH-3 in Homberger CVRPTW dataset with 600 cities. The best results appear in \textbf{bold}.}
\centering
\footnotesize
\begin{tabular}{lrrrrrrrrr} \toprule
\multirow{2}{*}{Instance group}             & \multicolumn{4}{c}{LKH-3}                                                                                         & \multicolumn{1}{c}{} & \multicolumn{4}{c}{\name-3}                                                                        \\ \cline{2-5} \cline{7-10}
 & \multicolumn{1}{r}{Best} & \multicolumn{1}{r}{Average} & \multicolumn{1}{r}{Trials} & \multicolumn{1}{r}{Time(s)} & \multicolumn{1}{r}{} & \multicolumn{1}{r}{Best} & \multicolumn{1}{r}{Average} & \multicolumn{1}{r}{Trials} & \multicolumn{1}{r}{Time(s)} \\  \hline
C1           & \textbf{14372.02}        & \textbf{14716.58}           & 10000.0                    & 202.41                      &                      & 14718.91                 & 14718.91                    & 10000.0                    & 63.10                       \\
C2           & \textbf{7786.52}         & 7958.47                     & 10000.0                    & 460.63                      &                      & 7874.96                  & \textbf{7874.96}            & 10000.0                    & 140.77                      \\
R1           & \textbf{19672.82}        & \textbf{19944.28}           & 10000.0                    & 528.05                      &                      & 20047.79                 & 20047.79                    & 10000.0                    & 283.47                      \\
R2           & \textbf{12685.21}        & \textbf{12811.52}           & 10000.0                    & 529.00                      &                      & 12841.78                 & 12841.78                    & 10000.0                    & 228.45                      \\
RC1          & \textbf{17116.87}        & \textbf{17321.11}           & 10000.0                    & 375.07                      &                      & 17423.80                 & 17423.80                    & 10000.0                    & 199.92                      \\
\bottomrule                    
\end{tabular}
\label{cvrptw-hoberger-600}
\end{table*}

%% file: table-cvrptw-Homberger-800.tex
\begin{table*}[!t]
\caption{Comparison results of \name-3 and LKH-3 in Homberger CVRPTW dataset with 800 cities. The best results appear in \textbf{bold}.}
\centering
\footnotesize
\begin{tabular}{lrrrrrrrrr} \toprule
\multirow{2}{*}{Instance group}             & \multicolumn{4}{c}{LKH-3}                                                                                         & \multicolumn{1}{c}{} & \multicolumn{4}{c}{\name-3}                                                                        \\ \cline{2-5} \cline{7-10}
 & \multicolumn{1}{r}{Best} & \multicolumn{1}{r}{Average} & \multicolumn{1}{r}{Trials} & \multicolumn{1}{r}{Time(s)} & \multicolumn{1}{r}{} & \multicolumn{1}{r}{Best} & \multicolumn{1}{r}{Average} & \multicolumn{1}{r}{Trials} & \multicolumn{1}{r}{Time(s)} \\  \hline
C1           & \textbf{25732.92}        & 26223.82                    & 10000.0                    & 564.06                      &                      & 25741.84                 & \textbf{26108.69}           & 10000.0                    & 1973.44                     \\
C2           & 11933.98                 & 12239.60                    & 10000.0                    & 975.16                      &                      & \textbf{11850.71}        & \textbf{12105.22}           & 10000.0                    & 2612.39                     \\
R1           & \textbf{33145.64}        & \textbf{33476.10}           & 10000.0                    & 567.54                      &                      & 33146.57                 & 33578.15                    & 10000.0                    & 2879.25                     \\
R2           & 20419.33                 & 20592.39                    & 10000.0                    & 610.02                      &                      & \textbf{20305.63}        & \textbf{20548.30}           & 10000.0                    & 2691.45                     \\
RC1          & \textbf{35300.66}        & \textbf{36206.53}           & 10000.0                    & 1237.10                     &                      & 35701.44                 & 36839.68                    & 10000.0                    & 5309.88                     \\
\bottomrule                    
\end{tabular}
\label{cvrptw-hoberger-800}
\end{table*}

%% file: table-cvrptw-Homberger-1000.tex
\begin{table*}[!t]
\caption{Comparison results of \name-3 and LKH-3 in Homberger CVRPTW dataset with 1000 cities. The best results appear in \textbf{bold}.}
\centering
\footnotesize
\begin{tabular}{lrrrrrrrrr} \toprule
\multirow{2}{*}{Instance group}              & \multicolumn{4}{c}{LKH-3}                                                                                         & \multicolumn{1}{c}{} & \multicolumn{4}{c}{\name-3}                                                                        \\    \cline{2-5} \cline{7-10}
 & \multicolumn{1}{r}{Best} & \multicolumn{1}{r}{Average} & \multicolumn{1}{r}{Trials} & \multicolumn{1}{r}{Time(s)} & \multicolumn{1}{r}{} & \multicolumn{1}{r}{Best} & \multicolumn{1}{r}{Average} & \multicolumn{1}{r}{Trials} & \multicolumn{1}{r}{Time(s)} \\ \hline
C1            & 42738.73                 & 43287.67                    & 9332.5                     & 822.12                      &                      & \textbf{42462.66}        & \textbf{42989.17}           & 9482.5                     & 2743.72                     \\
C2            & 17758.27                 & 18317.37                    & 10000.0                    & 1558.08                     &                      & \textbf{17349.16}        & \textbf{18108.47}           & 10000.0                    & 4247.85                     \\
R1            & \textbf{53627.22}        & \textbf{54409.12}           & 10000.0                    & 1171.42                     &                      & 54363.26                 & 55244.00                    & 10000.0                    & 6284.46                     \\
R2            & 29915.21                 & 30202.56                    & 10000.0                    & 851.46                      &                      & \textbf{29832.04}        & \textbf{30106.46}           & 10000.0                    & 3888.93                     \\
RC1           & \textbf{47794.32}        & \textbf{48311.10}           & 10000.0                    & 831.06                      &                      & 47813.20                 & 48680.03                    & 10000.0                    & 4528.96                     \\
\bottomrule       
\end{tabular}
\label{cvrptw-hoberger-1000}
\end{table*}

%% file: 07-Con.tex
In this work, we proposed a novel \name algorithm to improve the classic LKH algorithm for a typical NP-hard problem, the Traveling Salesman Problem (TSP). \name employs backbone information, $\alpha$-value, and distance to jointly guide the edge selection and further adopts a multi-armed bandit (MAB) to help select a promising combination of the three component metrics. In addition, we extended the \name framework to LKH-3 and denoted the resulting algorithm as \name-3, testing two classical variant problems of TSP and Vehicle Routing Problem (VRP), Colored TSP (CTSP) and Capacitated VRP with Time Windows (CVRPTW), to evaluate the performance and generalization capability of our proposed method. 
Extensive experiments show that both LKH and LKH-3 can be significantly improved by using our methods, indicating that our methods provide a generic algorithm framework and suggest an efficient way of using learning-based methods to boost local search heuristics for routing problems.


Though backbone information and MAB are not new technologies, the combination of reinforcement learning methods and combinatorial optimization is actually a magic recipe. 
We also perform ablation studies to explain the reason why the \name algorithm is effective, and the framework of the algorithm is of great significance. We believe that our proposed framework can be useful to other routing algorithms that share similarities to the $k$-opt operation. In future work, we will further explore the potential of \name, extend the algorithm to other routing problems and apply our idea to other tasks for real-world engineering applications.